\documentclass[12pt]{article}
\usepackage{aaspp4}
\usepackage{flushrt}
\journalid{337}{15 January 1989}
\articleid{11}{14}

\def\etal{{et al.}}
\begin{document}

\title{Structure of the Draco Dwarf Spheroidal Galaxy}
\author{Slawomir Piatek}
\affil{Dept. of Physics, New Jersey Institute of Technology, Newark,
NJ 07102 \\
     E-mail address: piatek@physics.rutgers.edu}

\author{Carlton Pryor\altaffilmark{1}}
\affil{Dept. of Physics and Astronomy, Rutgers, the State University of
New Jersey, 136~Frelinghuysen Rd., Piscataway, NJ 08854--8019 \\
E-mail address: pryor@physics.rutgers.edu}

\author{Taft E.\ Armandroff}
\affil{National Optical Astronomy Observatory, P.O.~Box~26732, Tucson, AZ 85726 
\\
     E-mail address: tarmandroff@noao.edu}
\and

\author{Edward W.\ Olszewski\altaffilmark{1}}
\affil{Steward Observatory, University of Arizona,
    Tucson, AZ 85721 \\ Email address: eolszewski@as.arizona.edu}

\altaffiltext{1}{Visiting Astronomer, Kitt Peak National Observatory,
National Optical Astronomy Observatory, which is operated by the
Association of Universities for Research in Astronomy, Inc. (AURA)
under cooperative agreement with the National Science Foundation.}

\begin{abstract}
 
	This article studies the structure of the Draco dwarf
spheroidal galaxy with an emphasis on the question of whether the
spatial distribution of its stars has been affected by the tidal
interaction with the Milky Way, using R- and V-band CCD photometry for
eleven fields.  The article reports coordinates for the center, a
position angle of the major axis, and the ellipticity.  It also reports
the results of searches for asymmetries in the structure of Draco.
These results, and searches for a ``break'' in the radial profile and
for the presence of principal sequences of Draco in a color-magnitude
diagram for regions more than 50~arcmin from the center, yield no
evidence that tidal forces from the Milky Way have affected the
structure of Draco.

\end{abstract}
\keywords{galaxies: dwarf --- galaxies: individual (Draco) ---
galaxies: stellar content --- galaxies: structure --- galaxies: Local Group}

\section{Introduction} 
\label{intro}

	This article studies the structure of the Draco dwarf
spheroidal (dSph) galaxy.  The dSph galaxies are characterized by
small size, low luminosity, low surface brightness, and old to
intermediate age stellar populations.  The Local Group dSphs are
clustered around and appear to be gravitationally bound to the much
more luminous spiral galaxies (see the review by van den Bergh 2000).
Draco, together with at least eight other dSphs, is a satellite galaxy
of the Milky Way.  Draco is 80 $\pm$ 7~kpc from the Sun (Aparicio,
Carrera, \& Mart\'{i}nez-Delgado 2001, ACMD hereafter) and, although
its tangential velocity is not known, the measured radial velocity,
corrected for solar motion, of $-98$~km~s$^{-1}$ (Olszewski
\etal\ 1995, Armandroff \etal\ 1995) implies that Draco is currently
approaching the Milky Way.

	Even though the masses of dSphs are small in comparison to
those of luminous spiral and elliptical galaxies, many of their
mass-to-light ratios ($\cal{M}/\cal{L}$\,s hereafter) are very large
(Aaronson 1983, Mateo 1998).  For example, the $\cal{M}/\cal{L}_{\rm
V}$ of Draco is 90 assuming that mass follows light (Armandroff
\etal\ 1995) and 340 -- 610 using more realistic models (Kleyna
\etal\ 2001) --- this is the highest measured value among the Galactic
dSphs.  The presence of non-luminous, or dark, matter in Draco is the
most direct interpretation of its large $\cal{M}/\cal{L}$.  However,
several authors have proposed alternative explanations for the large
measured $\cal{M}/\cal{L}$\,s of dSphs in general and Draco in
particular.  These explanations invoke either a modification of the
laws of gravity (MOND, Milgrom 1983) or the dSph being far from virial
equilibrium due to its interaction with the Galactic tidal field.

	Kuhn \& Miller (1989) and Kuhn (1993) proposed that a resonance
between the orbital frequency and the frequency of internal collective
oscillation modes of a dSph drives the dSph far from virial
equilibrium.  In this picture, a dSph with a large measured
$\cal{M}/\cal{L}$ is a gravitationally unbound stellar system which
does not dissipate quickly because the stars are on Galactic orbits
that keep them together.  However, Sellwood \& Pryor (1998) show
through numerical simulations that only a stellar system with even
lower central concentration than those for the observed dSphs has
collective modes that are not strongly damped and even in such a system
these modes are not excited by coupling to the orbital motion.  Oh,
Lin, \& Aarseth (1995) modeled weak, but non-resonant, tidal
interactions between the Galaxy and a dSph that led to the formation of
tidal debris around the dSph but did not increase the velocity
dispersion and $\cal{M}/\cal{L}$ to the values measured for real
dSphs.

	Piatek \& Pryor (1995) examined whether a single strong tidal
shock that disrupts a dSph can produce the large velocity dispersion
and $\cal{M}/\cal{L}$ values measured for real dSphs.  This study
found that such tidal interactions do not increase the velocity
dispersion but instead they produce a large velocity gradient along
the major axis.  Kroupa (1997) studied models in which the tidal
debris from the dSph is aligned along the line of sight, in which case
the velocity gradient masquerades as a large velocity dispersion.
These models require that the dSphs with large measured
$\cal{M}/\cal{L}$\,s are on nearly radial orbits.  They also predict
vertical broadening of the principal sequences, such as the horizontal
branch, in the color-magnitude diagram (Klessen \& Kroupa 1998,
Klessen \& Zhao 2001).

	The presence of tidal debris around a dSph does not prove that
their measured $\cal{M}/\cal{L}$\,s have been raised by tidal
interactions.  For example, Grillmair \etal\ (1995) and Leon \etal\
(2000) detected tidal debris around globular clusters, which have
measured $\cal{M}/\cal{L}_{V}$\,s of 1 -- 3 (Pryor \& Meylan 1993) that
agree well with those expected from their stellar populations.
However, because the spatial extent and the surface density of tidal
debris depends on the mass distribution, intrinsic $\cal{M}/\cal{L}$,
and orbital elements of the dSph, along with the Galactic potential,
detecting and quantifying the tidal debris can yield information about
these quantities (\textit{e.g.}, Kuhn 1993, Moore 1996, Johnston
\etal\ 1999a, Johnston \etal\ 1999c).  In addition, deriving the amount
and distribution of dark matter in a dSph using the kinematics of a
tracer population requires measurements of the projected density
profile of that population.  Limits on the central density of dark
matter are particularly sensitive to the shape of the profile at large
radii (Pryor 1994).

	Irwin \& Hatzidimitriou (1995; IH hereafter; see the references
therein for earlier work) derived radial profiles and structural
parameters for most of the Galactic dSphs using star counts from
Palomar and UK Schmidt telescope plates.  IH determined limiting, or
tidal, radii for a dSph by fitting single-component, isotropic King
models to its profile.  They noted that in many cases the dSph has a
profile that is above the fitted King model at large radii, which they
interpret as evidence for ``extra-tidal'' stars.  Many subsequent
studies have interpreted the IH tidal radius as the boundary between
gravitationally bound and unbound populations.  However, the two-body
relaxation time of every dSph is longer than its age (Webbink 1985) and
so there is no reason that the dSph must resemble a King model.
Indeed, it is possible to find many equilibrium models that fit
perfectly almost any projected density and projected velocity
dispersion profile (Dejonghe \& Merritt 1992).

	Several groups have recently studied the structure of the
Draco dSph.  Smith, Kuhn, \& Hawley (1997) detected apparent stars of
Draco extending up to 3~degrees east of the center --- far beyond the
28.3~arcmin tidal radius determined by IH.  They interpret these stars
as a tidally unbound population.  In contrast, Piatek \etal (2001;
hereafter P01) found evidence for Draco stars beyond the IH tidal
boundary, though only weak evidence for stars with distances as large
as 1~degree.  Odenkirchen \etal (2001b; hereafter Od01) studied the
morphology of Draco using Sloan Digital Sky Survey data for a wide
region around the galaxy.  They derived a limiting (or tidal) radius
for Draco of 49.5~arcmin, which is 75\% greater than the IH value.
Thus, they argue that the stars of Draco detected by P01 are within
the tidal boundary and therefore are gravitationally bound to Draco.
The Od01 upper limit for the surface density of Draco stars beyond
their limiting radius is a factor of ten lower than the surface
density detected by Smith, Kuhn, \& Hawley (1997).  Od01 argue that
this detection resulted from an incorrectly estimated background.
Interestingly, they also note that an exponential model fits their
data better than a King model, so the actual existence of a limiting
radius is called into doubt.  Finally, ACMD derive a radial profile of
Draco that is in broad agreement with that of Od01.

	The large discrepancy in the values for the limiting radius of
Draco obtained by IH and Od01 underscores the large uncertainties in
this fitted parameter.  Since the dynamics of the tidal debris is
decoupled from the internal dynamics of the dSph, the surface density
profile of the tidal debris and that of the dSph should have different
spatial structure.  Johnston \etal\ (1999b) performed numerical
experiments which show that a robust indicator of tidal debris around a
dSph is the presence of an abrupt change in the local power-law index
of the radial surface density profile of a dSph -- a ``break'' in the
profile.  The presence of tidal
debris causes the surface density to decrease less steeply in the outer
regions of a dSph.  Note, however, that Kroupa (1997) argues that a
profile resembling that of a real dSph may be produced by an unbound
population of stars.  It seems, therefore, that the limiting radius of
a dSph is not a trustworthy representative of the tidal boundary of a
dSph.

	In this article we report the results of a study of the
structure of Draco based on the R- and V-band photometric data for
eleven fields in and around Draco.  We derive such structural
parameters as the center, position angle, and ellipticity.  In
addition, we search for tidal debris using methods based on
color-magnitude diagrams, the radial profile of the surface density,
and the shape of isopleths of a map of the surface density.  We compare
our results to those from the existing studies of Draco.

	Section~\ref{data} describes the data and its reduction.
Section~\ref{cmds} presents the color-magnitude diagrams (CMDs) for the
new N1 and S1 fields, describes the procedure of discriminating between
non-members and possible members of Draco based on location in the CMD
and image morphology, and compares the sample of possible members of
Draco from this paper to those from ACMD and Od01.
Section~\ref{nomodel} derives model-independent structural parameters
of Draco and a map of the projected density of the galaxy.  The section
also comments on the reality of asymmetries in the projected density
map.  Section~\ref{discussion} summarizes and discusses our main
results.

\section{Data Acquisition and Reduction} 
\label{data}

	The data consists of R- and V-band photometry of objects from
eleven fields which are located in and around the Draco
galaxy. Photometry for nine of these fields is the same as those that
P01 used in their study. The two additional fields, N1 and S1, are
adjacent to the central C0 field of P01 in the north and south
directions, respectively, and extend beyond the tidal boundary along
the minor axis.  These additional fields were imaged with
the KPNO 0.9-m telescope using the $2048\times 2048$ T2KA CCD chip
(see P01 for more details).  There is a 0.55~arcmin overlap between
the N1 and C0 fields, while there is a 1.85~arcmin gap between S1 and
C0.  The N1 and S1 fields were taken under non-photometric conditions
and so we obtained two tie fields centered roughly halfway between the
N1 and C0 and S1 and C0 fields.  Table~\ref{basicinfo} gives the basic
information about the N1 and S1 fields and their corresponding tie
fields.  This table supplements Table~1 in P01, which gives this
information for the other nine fields.  Columns (1) and (2) list the
name of the field and the date of data acquisition. The next four
columns give the coordinates of the center of a field, first in the
equatorial system and then in the Galactic system.  Columns (7) and
(8) list the total exposure times for the V and R bands, respectively,
and the last two columns list the average value and root-mean-square
(rms) scatter around the average for the full width at half-maximum
(FWHM) of the stellar images in the V- and R-band frames,
respectively.  The FWHMs were measured in the same way as those in
P01.

	The data for the N1 and S1 fields have a more variable FWHM
than the majority of the P01 fields because they were taken before the
installation of the two-element field flattener on the 0.9-m
telescope.  The N1 R-band data have the worst seeing of any of our
data.  The N1 V-band data was taken through clouds and it has less
total exposure time than our other V-band data.  We discuss the impact
of these deficiencies on our results in Sections~\ref{completeness}\ and \ref{selecting}.

	The instrumental magnitudes of objects in the N1 and S1 fields
come from the same method described in P01 -- DAOPHOT (Stetson 1987,
1992, 1994) photometry of combined frames.  The tie frames determine
the transformation between the instrumental and standard magnitudes.
Because the tie frames were also taken under non-photometric
conditions we use them to transform N1 and S1 to the same photometric
system as C0.  To do so, we use the following procedure.  1) Determine
aperture magnitudes for objects in the tie frame using a 6-pixel
aperture radius.  2) Match the objects that are common to the C0 and
tie frames.  3) Using these objects with their standard magnitudes
measured in C0, derive a photometric transformation which converts
aperture magnitudes to standard magnitudes in the tie frame.  4) Match
objects common to the science and tie frames.  5) Using these objects,
derive a photometric transformation which converts instrumental
magnitudes in the science frame to standard magnitudes.

	Tables~\ref{photometry}A and \ref{photometry}B give the
standard $R$- and $V$-band photometry for the N1 and S1 fields,
respectively.  The first five columns in the tables give the ID, $x$
and $y$ coordinates on the $R$ frame, and $\alpha ({\rm J2000.0})$ and
$\delta ({\rm J2000.0})$ for an object.  The equatorial coordinates
come from plate solutions based on positions for stars in the
USNO-A2.0 catalog (Monet \etal\ 1998) using a recipe developed by Paul
Harding (personal communication).  Columns (6) and (7) give the
$R$-band magnitude followed by its uncertainty, $\sigma_{R}$. Columns
(8) and (9) list the same for the $V$ band. The last two columns give
the average CHI and SHARP values (described in Section~\ref{selecting}).

\subsection{Comparison of Photometry From Overlapping Fields}

	The regions of overlap between the N1 and C0 fields and
between the S1 and SE1 fields allow a comparison of the magnitudes for
the objects in common.  Figure~\ref{n1c0}\textit{a} plots the
magnitude difference $\Delta R \equiv R_{\rm N1} - R_{\rm C0}$ \textit{vs.}\
$R_{\rm N1}$ for the 47 objects that are common to the N1 and C0 fields and
are matched to within a 0.69~arcsec radius -- which is equivalent to about
1 pixel.  Similarly, Figure~\ref{n1c0}\textit{b} plots the magnitude
difference for the $V$-band.

The unweighted mean of $\Delta R$ is $-0.061 \pm 0.018$, where the
uncertainty is estimated from the rms scatter around the mean.  This
calculation excludes the three brightest objects in the sample, one of
which is out of the plot to the left and all of which are saturated in
the N1 image; the two dimmest objects, which have large uncertainties;
and the two objects near $R = 20$ with positive $\Delta R$, whose $R$-
and $V$-band photometry suggest are RR~Lyrae variables.  The offset between
the zero-points of the two fields is likely due to the poor focus near
the edges of the C0 frame.  However, our calibration procedure ensures
that the average zero-point shift between the two fields is zero.

The unweighted mean of $\Delta V$ is $-0.083 \pm 0.026$.  This calculation
excludes the two brightest objects, one of which is out of the plot to
the left and both of which are saturated in the C0 frame; objects dimmer
than $V = 23$, which have large uncertainties; and the same two likely
RR~Lyrae variables excluded from the calculation of the mean $\Delta R$.
Visual inspection of Figure~\ref{n1c0}\textit{b} shows that the mean
$\Delta V$ for the objects brighter than about $V = 21$
is closer to zero than the mean value calculated above, arguing that
there is no large zero-point difference for the V-band photometry.

	Figure~\ref{s1se1}\textit{a} plots the magnitude difference
$\Delta R \equiv R_{\rm S1} - R_{\rm SE1}$ \textit{vs.}\ $R_{\rm S1}$
for the 115 objects common to the S1 and SE1 fields.  Similarly,
Figure~\ref{s1se1}\textit{b} plots the magnitude difference for the
$V$-band.  The unweighted mean $\Delta R$ is $0.058 \pm 0.013$.  This
calculation excludes the brightest object, which is saturated, and
objects fainter than $R = 23$, which have large uncertainties.  Again,
this difference is likely due to the variable focus across the S1 field.
Excluding the brightest object and objects fainter than $V = 23$, the
unweighted mean $\Delta V$ is $-0.027 \pm 0.031$.  The plots and mean
magnitude differences for both the $R$- and $V$-band photometry show
that the zero points of the S1 and SE1 fields are in acceptable agreement.

\subsection{Completeness}
\label{completeness}

	The variation of the photometric completeness with position
distorts the measured structure of a stellar system.  Weighting each
object by the inverse of the completeness estimated for its magnitude
and position will reduce the distortion of the structure.  We perform
numerical experiments to determine the level of completeness as a
function of magnitude and position within a field for all of our
fields.

	In an experiment, we add artificial stars to both the R- and
V-band frames using a grid which places a star at the same location on
the sky.  The spacing of the grid is eight pixels (5.5~arcsec) in both
directions, which results in adding about 65,000 artificial stars to
each field.  The addition of such a large number of artificial stars in
a single experiment does not affect the crowding of the artificial
stars because the stars are arranged on a grid and thus do not overlap
each other.  An artificial star has the point-spread function which was
derived for its frame during the photometric reduction process.

	We perform five completeness experiments for every field,
reducing the $R$- and $V$-band frames together in the same way as the
real data.  The artificial stars in a given experiment and frame have
the same magnitude and color, which are chosen from the isochrone
described in P01.  The magnitude and color pairs used in the five
experiments are ($R$, $V-R$): (21.1,0.406), (21.6,0.395), (22.1,0.368),
(22.6,0.288), and (23.1,0.257).  An artificial star is counted as
recovered if it is measured in both frames and its position is within
one-half of a pixel of its input position, irrespective of how the
recovered magnitude compares to the input magnitude.

The completeness must be averaged over a region that is larger than
the spacing of the grid of artificial stars and that is large enough
to contain a fair sample of the local surface density of objects.  In
addition, this region should be smaller than the size of the structure
in Draco.  We adopt a circular window with a radius of 200 pixels in
order to smooth out the effects of saturated stars and their charge
overflow columns, which create the largest regions with big
fluctuations in the surface density of objects.  Using a weighted
average where the weight varies with distance, $r$, from the center of
the window, as $(1-(r/200~\textrm{pixels})^2)^2$ yields a completeness
that changes smoothly with position.  The C0 field has the largest
number of saturated stars and the steepest density gradients.  Thus,
in this field only, the estimates of both the completeness and the
surface density exclude regions around some of the brightest stars.
See Section~\ref{surfaced} for more details.

	Placing the center of the window at points with a spacing of
roughly 40 pixels in both directions yields an array of average
completenesses.  There are five arrays per field, one for each
magnitude and color pair of artificial stars.  An object with $16.4 < R
< 22.6$ has a completeness calculated by bicubic spline interpolation
in position and linear interpolation in magnitude.  Objects brighter
than $R=16.4$ have a completeness of 1.0, whereas objects dimmer than
$R=22.6$ are excluded from the analysis.

Figure~\ref{complete} plots the ratio of the total number of recovered
artificial stars to the number added vs.\ magnitude for each of our
eleven fields.  The C0 and N1 fields are significantly shallower than
the others.  The lower completeness of C0 at $R \leq 22.1$ is due to
the larger number of bright, saturated stars in this field.  The C0,
N1, S1, and E1 fields, imaged before the installation of the field
flattener, show a larger and more systematic variation of the
completeness within a field than the rest of the fields.  The
completeness decreases with increasing distance from the center in the
C0 and E1 fields.  The difference in completeness between the center
and edge can be as large as 0.4 when the completeness is changing most
rapidly with magnitude (see Figure~\ref{complete}).  In contrast, N1
and S1 have a larger completeness at the edge than in the center.  The
largest differences are similar to those for C0 and E1.  Only within
small regions near two corners of field C0 does the completeness fall
below 0.5 at $R = 22.6$.

The completeness simulations show that the errors in the magnitude and
color of a recovered star become rapidly larger as the input magnitude
approaches the limiting magnitude of the data.  Due to this effect, a
Draco star can appear far from the principal sequence of Draco in the
CMD and, thus, not be counted as a member of the galaxy.  Likewise, a
field star or galaxy can be scattered close to the principal sequence
of Draco and be counted as a member.  Measuring this effect would
require artificial star experiments that added objects throughout the
CMD.  Such experiments are difficult to implement because they require
knowing the true distribution of stars in the CMD and so they have not
been done by any study of the structure of a dSph.  This effect is most
important for the weak C0 and N1 fields and
Sections~\ref{surfaced}\ and \ref{rprofile}\ discuss the impact of this
effect on our results.  Wide-field CCD cameras are now available on
many large telescopes, thus obtaining higher-quality data is a better
approach to this problem than performing more elaborate completeness
experiments.

\subsection{Tangent Plane Projection}

	Expressing the positions of objects in a single, standard
coordinate system simplifies the study of the structure of a stellar
system.  Therefore, we convert positions of objects measured in pixels
within each frame into offsets in arcminutes from a chosen center by
performing a tangent plane projection. The offsets are with respect to
equatorial position 17$^h$~20$^m$~18.66$^s$ and
57$^\circ~55^\prime~5.55^{\prime\prime}$ (J2000.0), which corresponds
to the center of gravity of Draco determined by IH.  In our standard
system, the $X$ offset increases eastward and the $Y$ offset increases
northward.

\section{Color-Magnitude Diagrams}
\label{cmds}

	The top panel of Figure~\ref{cmdn1s1} shows a CMD for all
objects in the N1 field.  The bottom panel is the same for the S1
field. The magnitudes and colors of the objects in both panels are not
corrected for either redenning or extinction.

	The CMDs for the N1 and S1 fields show blue stars on the
horizontal branch (HB) of Draco near $R = 20$ and the lower
red-giant branch (RGB) stars at about $V-R = 0.3$ and $20 \lesssim R
\lesssim 23$.  This is expected since part of both fields are within
the tidal boundary of Draco.

\subsection{Reddening and Extinction}

The variation of extinction within and between our fields can also
distort the measured structure of a stellar system.  To reduce the
impact of this variation we determine average reddening and extinctions
for entire fields using the prescription of Schlegel~\etal\ (1998).  The
reddenings for the N1 and S1 fields are $0.030$ and $0.029$,
respectively.  These yield $V$- and $R$-band extinctions of 0.099 and
0.080, respectively, for the N1 field and 0.097 and 0.078,
respectively, for the S1 field.  Even after correcting for reddening
and extinction, small differences in our photometric zero points
remain, as evidenced by differences in the color of the blue edge of
the distribution of field stars in the CMD, measured as described in
P01.  This color is $0.234$ for the N1 field and $0.242$ for the S1
field.

\subsection{Selecting Samples Using CMD, SHARP, and CHI Criteria}
\label{selecting}

	The objects in our fields consist of stars and galaxies.  The
former are both Galactic foreground stars and members of Draco.  The
structure of Draco can best be measured using a sample of objects
containing a large number of Draco stars while at the same time
containing the smallest number of field stars and galaxies.  The lines
in the left panel of Figure~\ref{filters} outline the region in the CMD
where Draco stars are present.  The selection of this region is
discussed below.  The lines in the right panel outline the region where
stars are found in the diagram that plots SHARP value vs.\ magnitude.
A large positive value of the SHARP index calculated by the DAOPHOT
package indicates an object that is more extended than the point-spread
function -- which is likely a galaxy.  DAOPHOT also calculates CHI, a
measure of the quality of the point-spread function fit.  Values larger
than 5.0 occur for galaxies and spurious objects such as pixels with
charge overflow.  The remainder of this paper uses the sample of
objects that likely are stars on the basis of their CHI and SHARP
values and separates these into samples of likely members (the ``in''
sample) and likely non-members (the ``out'' sample) of Draco on the
basis of their position in the CMD.

The panels of Figure~\ref{filters} plot every second object from the C0
field and every tenth object from the other fields to show clearly both
the Draco sequences and the distribution of field stars and galaxies.
The width of the region outlining the RGB of Draco in the left panel of
Figure~\ref{filters} is proportional to the uncertainty in color for $R
< 21.6$.  The width is about $\pm$2$\sigma_{V-R}$, which is narrower
than the $\pm$3$\sigma_{V-R}$ used in P01.  The narrower width
increases the fraction of Draco members in the outlined region.  For $R
> 21.6$ the outlined region does not extend farther to the red with the
increasing uncertainty in color to avoid the large number of field
stars.  However, the region extends even farther to the blue to include
possible blue stragglers of Draco.  The region of allowed SHARP values
in the right panel of Figure~\ref{filters} is 40\% wider than that in
P01 to accommodate the more variable focus of the C0, N1, S1, and E1
fields.

	The artificial star simulations described in
Section~\ref{completeness} show that the large photometric
uncertainties of the faint objects will scatter them both into and out
of the ``in'' and ``out'' samples.  This effect can alter the measured
structure of Draco because the distribution of objects in the CMD is
not the same in all of our fields.  For example, our C0 and N1 fields
are shallower than the others.  As discussed in
Section~\ref{completeness}, we do correct our data for completeness but
we have chosen not to correct for this scattering effect.

\subsection{Background Surface Density}
\label{background}

	We determine the background surface density of the ``in'' and
``out'' samples using the most distant fields:  N2, E2, S2, SW2, and
W2, which are about 1~degree from the center of Draco (see Figure~1 in
P01).  The weighted mean surface density of the background for the
``in'' sample is $1.446\pm 0.026$~arcmin$^{-2}$.  The weight for each
surface density is the inverse of the square of the sampling
uncertainty.  The $\chi^2$ of the scatter around the mean is 4.96 for
three degrees of freedom (the S2 and SW2 fields were combined into one
since they overlap).  The probability of exceeding this value by chance
is 0.17, so there is no evidence for more variability than expected by
counting statistics and the stated uncertainty in the mean is
realistic.  For the ``out'' sample, the $\chi^2$ of the scatter around
the weighted mean is 22.8 and the probability of exceeding this by
chance is $4.4\times 10^{-5}$.  The clustering of galaxies causes the
larger variability of the surface density compared to that expected
from the sample size (see the detailed discussion in P01).  Thus, for
the ``out'' sample we adopt an unweighted average background of
$3.91\pm 0.12$~arcmin$^{-2}$, where the uncertainty is based on the
scatter around the mean.

\subsection{Comparison with Other Samples}
\label{goodsample}

The current paper is one of several that have recently studied the
structure of Draco using star counts.  This section briefly compares
our ``in'' sample with the corresponding samples from ACMD and Od01.
The goal is to assess how well these samples determine the structure
of Draco.

The ``in'' sample in this paper is intermediate in both limiting
magnitude and radial exent compared to those in ACMD and Od01.  Its
limiting magnitude is about 2.6 magnitudes below the level of the
horizontal branch compared to about 4.0 magnitudes below for ACMD and
1.8 magnitudes below for Od01.  The Od01 sample has complete azimuthal
coverage to a distance of 2~degrees from the center of Draco whereas
the ``in'' sample has six inner fields extending to about 50~arcmin
from the center and five background fields about 1~degree from the
center (see Figure~1 in P01).  The AMCD sample has three fields
extending to about 50~arcmin from the center of Draco.

The central surface density and its ratio to the background surface
density far from the center are two of the best measures of the quality
of a sample for determining the structure of Draco.  The central
surface density of the ``in'' sample (presented in
Section~\ref{rprofile}) is half that of the ACMD main-sequence sample
but it is about three times that of the Od01 sample.  The ratio of the
central and background surface densities for the ``in'' sample is 3--5
times smaller than those of the ACMD and Od01 samples, reflecting the
better discrimination against non-members in the CMD allowed by the
homogenous photometry in Od01 and by the deeper photometry in ACMD.
These comparisons show that the ``in'' sample is better than the Od01
sample for studying the inner regions of Draco.  The ACMD sample would be
better still, however this work is focused on the stellar populations
rather than on the structure of Draco.

\section{Model-independent Structure}
\label{nomodel}

This section examines the structure of Draco without assuming a
parametric model.  It first discusses the derivation of a smooth
surface density map for objects from our data using an adaptive kernel.
Discussions of a contour plot of the surface density map and of estimates
of the center of the galaxy and of the position angle of its major axis
follow.  The section ends by discussing the centers, position angles,
and ellipticities resulting from fitting ellipses to the smooth surface
density map.

\subsection{Adaptive Kernel Estimate of the Smooth Surface Density}
\label{adaptive}

The construction of a surface density map from the positions of objects
on the sky requires smoothing.  Kernel estimators are commonly used for
this purpose (e.g., Silverman 1986).  For systems with large variations
in the surface density a kernel whose width decreases as the density
increases -- an adaptive kernel -- recovers the maximum amount of
information.

We use a parabolic kernel to construct the smoothed density map for
Draco on a grid of points following the methods outlined in Silverman
(1986).  A kernel with a fixed width creates a ``pilot'' density
estimate.  The final surface density map uses an adaptive kernel whose
width is inversely proportional to the square root of the pilot density
at each grid point.  This procedure keeps the number of objects
contributing to each kernel area approximately constant.

The construction of the surface density map from our data is
complicated by the incomplete areal coverage of our fields.  There are
swaths of missing data north-west, south-west, and north-east from the
galaxy and small gaps in the data between the CO and W1 and C0 and S1
fields.  There are also numerous holes in the data due to bright and
saturated stars.  We generate artificial data to fill in the empty
regions within an area of 120~arcmin by 120~arcmin centered on Draco.
Most of the empty regions, shown in white in Figure~\ref{dcontour},
have artificial objects generated from a constant distribution function
scaled to have the average surface density of the background fields.
The artificial objects in grey regions in the figure are drawn from the
adjacent regions with the same area and containing real objects.  While
filling holes in the data is required for kernel estimates of the
surface density, the surface density within a kernel width of large
regions of artificial data should be treated with caution.

\notetoeditor{We request that Figure 6 be two columns wide in the
published paper for legibility.}

\subsection{Contour Plot of the Surface Density}
\label{surfaced}

	Figure~\ref{dcontour} shows contours of constant surface
density for Draco.  The kernel size ($w$ in the kernel $1-(r/w)^2$,
where $r$ is the radial distance from the object) for the pilot
estimate is 3.1~arcmin, producing largest and smallest adaptive widths
of 1.5~arcmin and 9.9~arcmin, respectively.  Approximately 90 objects
are within the area of the adaptive kernel, yielding a fractional
uncertainty in the density of 0.11.  The solid contour lines represent
values of the surface density above the adopted background density
whereas dashed lines represent values below.  Contours are drawn at
0, $\pm$1, $\pm$2, 3, 4, and 5$\sigma$ from the background level, where
$\sigma$ is the fractional uncertainty in the density estimate times
the background density.  The contour levels at higher surface density
are spaced by 3$\sigma$, where now $\sigma$ is the fractional
uncertainty times the surface density at the previous contour.  The
actual contour levels are in the figure caption.

	The position angle of the major axis of the innermost contour
is different from those for the more distant contours.  We show in
Section~\ref{statexp} that the statistical significance of this apparent
difference is very low, arguing that this difference is caused by the
sampling uncertainty in the surface density estimate.

	The contours also show an apparent lopsidedness: there seems to
be a shoulder about 10~arcmin to the east of the center and a steeper
gradient on the north side of the galaxy than on the south side.  We
test for the statistical significance of asymmetries along the major
and minor axes in Section~\ref{asymmetry}.  We find that these
asymmetries are either due to the sampling uncertainty (along the major
axis) or problems with the photometry (along the minor axis).

	Figure~\ref{dcontour} shows that surface densities above
background extend beyond 28.3~arcmin -- the tidal radius determined by
IH -- along and close to the major axis, which confirms the results of
IH, Od01, and ACMD that Draco extends beyond the IH tidal boundary.
IH argued on the basis of an abrupt change in the slope of the outer
parts of their radial profile that these stars are not gravitationally
bound, whereas the other authors argued on the basis of the absence of
such a change that they are bound.  A map of surface density can
reveal tidal debris if the outermost contours show an S-shape
distortion or extended tidal tails.  Grillmair \etal\ (1995), Leon
\etal\ (2000), and Odenkirchen \etal\ (2001a) have detected such
features around Galactic globular clusters.  The visibility of
S-shaped distortions around a Galactic dSph is likely to be suppressed
by our nearly-in-the-orbital-plane position with respect to the dSph.
However, the line-of-sight projection of this debris could still
introduce irregularities into the structure.  The contours in
Figure~\ref{dcontour} do not show any obvious signs of tidal
distortions or extended tidal tails.  However, our data do not extend
far enough to rule out the existence of tidaly induced distortions in
Draco.

	The surface density is below background about 20~arcmin to the
north and about 25~arcmin to the southeast of the galaxy at the
1--2$\sigma$ level.  Such regions are expected from fluctuations caused
by counting statistics.  However, the surface density is also lower
than background in these regions in a similar map constructed using the
``out'' sample.  The surface density of objects classified as
non-stellar in the Sloan Digital Sky Survey data in the vicinity of
Draco (York \etal\ 2000) also show lower than average values in these regions.
All of this evidence suggest that large-scale structure in the
distribution of galaxies, which are not completely eliminated from our
``in'' sample, is at least partly responsible for the low surface
densities.  The surface density of galaxies is lower in the southeast
region than in the north region, whereas Figure~\ref{dcontour} shows
the reverse.  We think that the large photometric uncertainties in the
N1 field compared to our other fields have scattered more stars from
the ``in'' sample into the ``out'' sample, thus reducing the measured
surface density (see the discussion in Sections~\ref{completeness}\ and
\ref{selecting}).  This spuriously low surface density is likely responsible
for the steeper density gradient seen on the north side of Draco compared
to the south side.

\subsection{Center and Position Angle}
\label{xyp}

If the distribution of stars in Draco were not symmetric, then the
center of Draco would depend on the method used to measure it.  The
first two methods employed in this study use a minimum of assumptions
to find the center of Draco from the ``in'' sample of stars in the C0
field.

	The first method finds a center of symmetry for Draco using a
mirrored autocorrelation of the one-dimensional distribution of stars
in either the $X$ or $Y$ direction calculated in a sliding window
20~arcmin wide in $X$ and 14~arcmin wide in $Y$.  This method yields a
center at $X=-0.28$~arcmin $Y=0.11$~arcmin.  Varying the size of the
sliding window implies that the uncertainties in these positions are
on the order of 0.1~arcmin.

	The second method finds a center that minimizes the fractional
rms scatter in the number of stars in the four quadrants of a circular
aperture about their mean.  The radius of the aperture equals the
shortest distance between its center and the nearest edge of the C0
field.  This method aligns the boundaries of the four quadrants with
the major and minor axes of Draco, thus it also yields the position
angle of the major axis.  The resulting center is $X=0.16\pm
0.18$~arcmin and $Y=0.15\pm0.12$~arcmin and the position angle of the
major axis is $90.6\pm 4.6$~degrees. The uncertainties come from one
thousand bootstrap determinations of the center and position angle.
Each determination draws a sample of the same size as the original
sample from the original sample with replacement.

\subsection{Do ellipticity, Position Angle, and Center Depend on
Semi-major Axis?}
\label{statexp}

With the assumption that the contours of constant projected density are
ellipses, fitting ellipses gives another determination of the center
and position angle of the major axis along with the ellipticity.  The
variation of these quantities with the length of the semimajor axis is
an indication of the presence of asymmetries.  The panels of
Figure~\ref{epxcyc} show the dependence of the ellipticity ($e$),
position angle (PA), $X$-coordinate of the center ($X_c$), and
$Y$-coordinate of the center ($Y_{c}$) on the semimajor axes of
ellipses ($a$) fit to the estimate of the surface density described in
Section~\ref{surfaced} and shown in Figure~\ref{dcontour}.  The best
fit ellipse has the minimum rms fluctuation in the surface density
measured at 360 points equally spaced in arclength around the ellipse.
Each triangle in Figure~\ref{epxcyc} is the value of the structural
parameter determined from the fit to the actual sample.  The
corresponding square and its associated error bar are the mean value of
the structural parameter determined from 1000 bootstrap simulations and
the rms scatter around this mean, respectively.  The smallest value of
the semimajor axis fitted is set by the minimum kernel width and the
largest by the presence of gaps in the data.

	Panel~a) of Figure~\ref{epxcyc} shows that the ellipticity,
$e$, does not vary significantly with semimajor axis. The weighted
average value of $e$ is $0.331\pm0.015$ and the total $\chi^{2}$ is
1.4.  The uncertainties used to calculate these quantities come from
the bootstrap experiments.  Approximately only every other point is
independent because of the smoothing.  However, the $\chi^2$ per the
smaller true number of degrees of freedom is still less than one, which
shows that the variation in $e$ is not significant.

This conclusion is strengthened by results from Monte-Carlo simulations
of fitting ellipses to the density field of a symmetric model of
Draco.  The model is the power law with a core used by Kleyna
\etal\ (1998) and has $e=0.29$, PA = 88\arcdeg, $X_c = Y_c = 0$~arcmin, a core
radius of 17.7~arcmin, and a power-law exponent of 4.1.  The values for
the last two parameters come from fits to our data.  Each
realization of the model is smoothed adaptively in the same way as the
real data.  The panels in Figure~\ref{epxcycmc} show the mean values of
$e$, PA, $X_c$, and $Y_c$ from 1000 Monte Carlo experiments.  The error
bar is the rms scatter around the mean.  Panel a) of
Figure~\ref{epxcycmc} shows that the mean $e$ is biased upwards to
about 0.4 for a semimajor axis of 3.0~arcmin.  Our fitted ellipticities
for the real data show a very similar trend.

	Panel b) of Figure~\ref{epxcyc} plots the position angle of
the major axis, PA, \textit{vs.}\ semimajor axis.  The PA varies from
$52^{\circ}\pm17^{\circ}$ at $a=3.0$~arcmin to
$90.0^{\circ}\pm2.2^{\circ}$ at $a=12.0$~arcmin.  The innermost point
is about two standard deviations below the mean of the points at larger
semimajor axes, implying a marginally significant change of PA.  The
orientation of the surface density contours in Figure~\ref{dcontour}
also shows this change.  Panel b) of Figure~\ref{epxcycmc} shows that
the rms scatter in the PA at a semimajor axis of 3.0~arcmin is
large enough that statistical fluctuations can explain the change
observed in the real data.

	Panel c) of Figure~\ref{epxcyc} shows that $X_{c}$
systematically increases with increasing semimajor axis.  Panels c) and
d) of Figure~\ref{epxcycmc} show that a symmetric model produces no
such trends, on average.  The trend in the real data reflects a small
asymmetry of the contours in Figure~\ref{dcontour} -- as described in
Section~\ref{surfaced}.  Section~\ref{asymmetry} tests the statistical
significance of this apparent asymmetry and finds that it is not.  The
$\chi^2$ for the $X_{c}$ values about their weighted mean of $-0.10 \pm
0.11$~arcmin is about one per degree of freedom, which also argues that
the trend is not significant.

	Finally, panel d) of Figure~\ref{epxcyc} plots the
$Y$-coordinate of the center, $Y_{c}$.  There is no evidence for a
dependence of $Y_{c}$ on semimajor axis. The weighted mean of $Y_{c}$
is $0.13\pm0.06$.

\subsection{Asymmetry}
\label{asymmetry}

	The contour plot of Draco depicted in Figure~\ref{dcontour}
shows an apparent ``shoulder'' about 10~arcmin to the east of the
center, approximately along the major axis, and a steeper gradient
beyond 10~arcmin from the center on the north side than on the south,
approximately along the minor axis.  The statistical significance of an
asymmetry can be ascertained by measuring how often asymmetry can arise
by chance from a symmetric model of Draco due to the finite size of the
sample of stars.

Kleyna \etal (1998) defined an asymmetry statistic, $\beta =
(d_1/d_2) - 1$, where $d_1$ and $d_2$ are the distances along the
major or minor axis between the point of highest surface density and
the points where the surface density has fallen by a factor of two and
$d_1 > d_2$.  A system is symmetric if $\beta=0$ and it is asymmetric
if $\beta > 0$.  As defined, the $\beta$ statistic can measure the
asymmetry observed in our surface density map along the major axis.  It
cannot measure the observed asymmetry seen at a larger distance along
the minor axis.  However, the latter asymmetry is likely due to
problems with the photometry in the N1 field discussed in
Sections~\ref{completeness}\ and \ref{selecting}.  Thus, we do not
investigate this asymmetry along the minor axis.

Measuring $\beta$ for the major axis using the same projected density
depicted in Figure~\ref{dcontour} and a major axis position angle of
88~degrees yields 0.24.  Similarly, the value of $\beta$ for the minor
axis is 0.47.   Monte Carlo simulations using the symmetric model
described in the previous section give a larger value for the $\beta$
along the major axis 81\% of the time and for the $\beta$ along the
minor axis 69\% of the time.  Thus, the data shows no evidence that
Draco is asymmetric along either its major or minor axis.

\subsection{Radial Profile}
\label{rprofile}

	The top panel of Figure~\ref{starcount} plots the
completeness-corrected surface density of objects from the ``in''
sample calculated in elliptical annuli \textit{vs.}\ the semi-major
axes of the annuli.  We calculate the area of the overlap between each
annulus and the data by dividing the sky into 2.8~arcsec square bins
and summing the areas of the bins whose centers are within the annulus
and within the boundaries of our fields.  The center of the annuli is
at $X = -0.10$~arcmin and $Y = 0.13$~arcmin.  The annuli have an
ellipticity of 0.33 and a major-axis position angle of 91$^\circ$.  The
semi-major axis of an annulus is the average of the semi-major axes of
the ellipses passing through the objects in the annulus.  Each point
has an error bar equal to the surface density divided by $\sqrt{N}$,
where $N$ is the actual (not completeness corrected) number of objects
in the annulus.  The bottom panel of Figure~\ref{starcount} plots the
background-subtracted surface density using the value of
1.446~arcmin$^{-2}$ from Section~\ref{background}.  Here, the error bar
includes the uncertainty in the background, 0.026~arcmin$^{-2}$, added
in quadrature.  The horizontal dashed line in both plots represents the
surface density of the background.  Table~\ref{radialp} tabulates the
surface densities shown in the top panel and their radii.

The top panel of Figure~\ref{starcount} shows that the surface density
profile is approximately constant within a semi-major axis of 5~arcmin
and decreases beyond.  The profile flattens as it approaches the
background surface density, however it never becomes flat and continues
to decrease to values below the background found from the fields at
larger radii.  The surface densities that are below background in the
outermost four annuli are likely caused in part by inadequate
corrections for incompleteness, as discussed in
Sections~\ref{completeness} and \ref{selecting}, and in part by the
statistical accident that the portions of these annuli sampled by our
data are regions of true low surface density (see
Section~\ref{surfaced}).
 
Figure~\ref{allrp} shows the background-subtracted projected density
profiles of Draco from IH, Od01, and this article.  The open squares
are the IH profile taken from their Table~3, with the first 24 points
in the table binned by two and the remainder by four.  The error bars
are based on counting statistics and the number of stars in each binned
point.  The triangles are the Od01 profile, where the open symbols are
their S1 sample and the solid symbols are their S2 sample.  The solid
squares are the profile in the bottom panel of Figure~\ref{starcount}.
The error bars for all of the profiles include the uncertainty in the
background added in quadrature.  The vertical normalizations of the
four profiles make the innermost point of each equal to one.

All four profiles in Figure~\ref{allrp} agree well up to a radius of
about 20~arcmin.  IH interpreted the apparent flattening of their
profile beyond 20~arcmin as evidence for extra-tidal stars.  However,
the IH profile has roughly as many points below background as above in
this region, implying that the flattening is not statistically
significant (Od01).  In addition, the IH sample does not exclude
galaxies and so the error bars shown are too small because the
fluctuations in the surface densities of galaxies exceed those from
Poisson statistics.  Therefore, we think that the IH profile is
consistent with those of Od01 and this article.  The Od01 profile and
that from this article agree well out to a semi-major axis of about
40~arcmin and neither one shows an abrupt change of slope (a
``break'').  We conclude that none of the profiles shows unambiguous
evidence for tidal debris around Draco.

\subsection{CMD Outside of the Tidal Boundary}

	Od01 report a tidal radius for Draco of 49.5~arcmin, based on
fitting a King (1966) model, and ACMD report a tidal radius of
42~arcmin, based on a visual inspection of their radial surface
density profile.  The difference between these two values reflects the
difficulty of measuring the tidal radius.  However, both studies
demonstrate convincingly that Draco extends beyond the 28.3~arcmin
tidal radius found by IH.  Od01 argue that this more extended profile
explains the stars of Draco detected by P01 beyond the IH tidal
boundary.  To search for Draco stars at still larger radii, we plot a
CMD for the objects inside and outside of the Od01 tidal boundary --
which is an ellipse with semimajor axis of $49.5$~arcmin, a position
angle of $88^{\circ}$, and centered at $\alpha$ = 17$^{\rm h}$
20$^{\rm m}$ 13.2$^{\rm s}$ and $\delta$ = 57$^\circ$ 54$^{\prime}$
54$^{\prime\prime}$ (J2000), which corresponds to
($-0.73^{\prime}$,$-0.19^{\prime}$) in our standard coordinate system.

	The top panel in Figure~\ref{cmd-tb} is a CMD for the objects
located inside of the Od01 tidal boundary and the bottom panel is the
corresponding plot for the objects outside.  The CMD in the bottom
panel shows no clear visual evidence of the principal sequences of
Draco.  This lack of visual evidence does not necessarily imply an
absence of Draco stars beyond the Od01 tidal boundary.  There could
simply be too few Draco stars present to be noticeable in the CMD.
Indeed, such a population is expected if the profile of Draco is an
exponential, which has no limiting radius.  Od01 and ACMD find that an
exponential profile is a good fit to their projected density profiles.

\section{Summary and Discussion}
\label{discussion}

	The average center of Draco measured with three methods
described in Sections~\ref{xyp}\ and \ref{statexp}\ is at $\alpha$ =
17$^{\rm h}$ 20$^{\rm m}$ 18.1$^{\rm s}$ and $\delta$ = 57$^\circ$
55$^{\prime}$ 13$^{\prime\prime}$ (J2000).  The uncertainty is about
0.1~arcmin in both coordinates.  This center is 40~arcsec east and 19~arcsec
north of the center reported by Od01, a difference that is somewhat
larger than that expected from the 25~arcsec uncertainty in right
ascension and 11~arcsec uncertainty in declination of the Od01 value.
However, the difference is smaller than twice the uncertainty and,
thus, not statistically significant.

	The position angle of the major axis of Draco is $90.6\pm
4.6$~degrees based on the objects within approximately 10~arcmin of the
center (see Section~\ref{xyp}).   This value is in agreement with the
$88\pm 3$~degrees measured by Od01.  Fitting ellipses to the smoothed
surface density, described in Section~\ref{statexp}, gives a range of
position angles consistent with the above values.  Od01 and this study
find no evidence that the position angle of the major axis varies with
semi-major axis.

	The ellipticity of Draco, determined from the average of the
values for the ellipses fitted to the smoothed surface density, is
$0.331\pm0.015$.  This average value is greater than the $0.29\pm
0.02$ determined by Od01.  Our value is less reliable because of the
problems with the photometry and completeness corrections in the N1
field described in Sections~\ref{completeness} and \ref{selecting}.

Tidal debris projected onto a bound dSph can produce a small asymmetry
in the surface density map (Mayer \etal\ 2001).  A larger asymmetry
might arise if the observed dSph consists primarily of unbound tidal
debris (Kroupa 1997, Klessen \& Kroupa 1998).  The contours of the
smoothed surface density in Figure~\ref{dcontour} show an apparent
``shoulder'' about 10~arcmin east of the center.  We tested the
statistical significance of the apparent asymmetry along the major and
minor axes and found that both can occur by chance 81\% and 69\% of
the time, respectively.  Therefore, we find no compelling evidence for
asymmetries in Draco, tidally induced or otherwise.

Figure~\ref{allrp} shows that the radial profile of Draco from this
study agrees with the radial profiles from IH and Od01 within the
uncertainties.  The radial profile of Draco does not show evidence of
an abrupt change, or break, in the slope. In addition, the cmd in
Figure~\ref{cmd-tb} does not show the principal sequences of Draco for
the region beyond the Od01 tidal boundary, which is also about the
last point in our radial profile.  Thus, we find no evidence that
Draco is surrounded by tidal debris.

\acknowledgments

CP and SP acknowledge support from the National Science Foundation through
the grant AST 00-98650, while EWO acknowledges support through the grants
AST 96-19524 and AST 00-98435.  This research has made use of the
Astronomical Data Center (ADC) at the NASA Goddard Space Flight
Center.

\clearpage
\begin{center}
Figure Captions
\end{center}

Fig. 1. (a) A comparison of the R-band magnitudes for the 47 objects
common to the N1 and C0 fields. (b) The same as Fig.~1a for the V
band.

Fig. 2. (a) A comparison of the R-band magnitudes for the 115 objects
common to the S1 and SE1 fields.  (b) The same as Fig.~2a for the V
band.

Fig. 3. Average completeness as a function of magnitude for the eleven
fields.  Each point is the ratio of the number of recovered to the number
of added artificial stars.

Fig. 4. Color-magnitude diagrams for N1 (top panel) and S1 (bottom
panel). Only objects with CHI values less than 5 and SHARP values
within the limits shown in Fig.~5 appear in these diagrams.

Fig. 5.  Left panel: A color-magnitude diagram showing every second
object from the C0 field and every tenth object from the other ten
fields.  The solid contours outline the principal sequences of Draco.
Right panel: R-band magnitude \textit{vs.} SHARP for the same sample of
objects as that in the left panel. The solid lines represent the SHARP
limits we adopt to discriminate between stars and galaxies. The objects
within the lines are likely to be stars and those outside are likely to
be galaxies or spurious objects. The two panels define the ``in'' and
``out'' samples: the ``in'' sample consists of objects within the
outlined regions in both panels, whereas the ``out'' sample consists of
those that are outside of the outlined region in the left panel (and
brighter than $R=22.6$) but within the outlined region in the right
panel.

Fig. 6.  Isopleths for a smoothed surface density map of Draco using
the ``in'' sample of objects.  Dashed contours correspond to values
below the background.  The lowest contour levels are 0, $\pm 1$, $\pm
2$, 3, 4, and 5$\sigma$ from the background, where $\sigma$ is the
uncertainty of the smoothed density for locations in the map with
densities near background.  The spacing between the levels of the
higher contours are $3\sigma$, where $\sigma$ now is the uncertainty in
the density at the value of the lower contour.  The contour levels from
the lowest to the highest are:  1.145, 1.295, 1.446, 1.597, 1.747,
1.898, 2.199, 2.886, 3.787, 4.970, 6.522, 8.560, 11.234, 14.743,
19.348, 25.391~arcmin$^{-2}$.  The grey and white areas contain no
data, see the text for details.

Fig. 7.  The dependence on semi-major axis, a, of ellipticity (panel
a), position angle of the major axis (panel b), x-coordinate of the
center (panel c), and y-coordinate of the center (panel d) for ellipses
fitted to the smoothed density map of Draco.  An open triangle is the
value of the structural parameter determined from fitting to the map
shown in Figure~6.  An open square and its associated error bar are the
mean value of the structural parameter and the rms scatter around the
mean, respectively, determined from 1000 bootstrap experiments.  Only
approximately every other point is independent because of the smoothing
in the density map.

Fig. 8.  The dependence on semi-major axis, a, of ellipticity (panel
a), position angle of the major axis (panel b), x-coordinate of the
center (panel c), and y-coordinate of the center (panel d) for ellipses
fitted to the smoothed density map for a symmetric model of Draco.
Each point and its associated error bar are the mean and rms scatter
around the mean, respectively, determined from 1000 Monte Carlo
experiments.

Fig. 9.  Radial profile of Draco before (top panel) and after (bottom
panel) subtracting the background. Table~3 tabulates the values of the
surface density shown in the top panel, their uncertainties, and the
radii.

Fig. 10.  Radial profiles after subtracting the background from Irwin
\& Hatzidimitriou (1995; IH) (open squares), Odenkirchen \etal\ (2001b;
Od01) (open and solid triangles for samples S1 and S2, respectively),
and this study (solid squares). The normalization of each profile
makes the surface density of the innermost point equal to one.

Fig. 11.  Color-magnitude diagram for the objects inside (top panel)
and outside (bottom panel) of the Od01 tidal boundary of Draco.  This
boundary is an ellipse centered at $X=-0.73$~arcmin and
$Y=-0.19$~arcmin with an ellipticity, position angle of the major axis,
and a semi-major axis of 0.30, $88^{\circ}$, and 49.5~arcmin,
respectively.

\setcounter{figure}{0}

\begin{figure}
\caption{}
\label{n1c0}
\end{figure}

\begin{figure}
\caption{}
\label{s1se1}
\end{figure}

\begin{figure}
\caption{}
\label{complete}
\end{figure}

\begin{figure}
\caption{}
\label{cmdn1s1}
\end{figure}

\begin{figure}
\caption{}
\label{filters}
\end{figure}

\begin{figure}
\caption{}
\label{dcontour}
\end{figure}

\begin{figure}
\caption{}
\label{epxcyc}
\end{figure}

\begin{figure}
\caption{}
\label{epxcycmc}
\end{figure}

\begin{figure}
\caption{}
\label{starcount}
\end{figure}

\begin{figure}
\caption{}
\label{allrp}
\end{figure}

\begin{figure}
\caption{}
\label{cmd-tb}
\end{figure}

\setcounter{table}{0}
\begin{table}
\caption{}
\label{basicinfo}
\end{table}

\begin{table}
\caption{}
\label{photometry}
\end{table}

\begin{table}
\caption{}
\label{radialp}
\end{table}
	
\end{document}